\documentclass[12pt]{article}
\usepackage{amssymb,amsmath,latexsym,graphicx, bm, mathrsfs, hyperref}
\usepackage{graphics,graphicx, cancel}
\usepackage[dvipsnames]{xcolor}

\usepackage[british]{datetime2}

\begin{document}

\title{Zermelo Wind: a geometrization \\ 
of the frame dragging effect 
\vspace{-0.25cm} 
\author{Sumanto Chanda \\ \vspace{-0.5cm} \\ 
\textit{Indian Institute of Astrophysics} \\ 
\textit{Block 2, 100 Feet Road, Koramangala,} 
\textit{Bengaluru 560034, India.} \\ 
\texttt{\small sumanto.chanda@iiap.res.in}}}

\maketitle

\thispagestyle{empty}
\vspace{-1.0cm}
\abstract{
In this article I discuss Zermelo's navigation problem 
in spacetime as a geometrization of the frame dragging 
effect, and recast various examples involving the latter 
into Zermelo form. 
I start by describing a stationary spacetime in 
Zermelo's form and show that the Zermelo wind 
is the drift velocity under frame dragging effect. 
Then we discuss various problems in this context, 
such as Hubble expansion of the universe and 
accelerated frames in special relativity. 
Another example I will discuss is the self-gravitating 
disk around a black hole in post-Newtonian (PN1) 
approximation to describe the anti-dragging effect 
in terms of a Zermelo wind. 
}

\vspace{-0.5cm}
\tableofcontents

\newpage 

\numberwithin{equation}{section}

\section{Introduction}
\setcounter{page}{1}

Zermelo's navigation problem was first proposed 
in 1931 by Ernst Zermelo \cite{zermelo} as a 
modification of the optimal travel time problem from 
variational calculus problem involving a wind function. 
The original problem discussed the question of the 
trajectory between two given points requiring minimal 
time for a ship moving with constant speed under the 
influence of a wind. 
When the Zermelo problem is formulated in spacetime, 
the wind effectively describes a locally moving background 
frame on top of motion occurs that must be taken into 
account when writing the metric. 
The problem was more recently generalized by Shen 
to study motion in Finsler space in 2003 \cite{shen}, 
also showing that the minimal-time trajectory describes 
a Randers-Finsler (RF) type of geodesic \cite{randers}.

Randers' type of Finsler geometry is a modification 
of Riemannian geometry by adding a linear term akin 
to a 1-form derived from a magnetic potential \cite{randers}. 
Geodesics described via such geometry often derives 
from stationary spacetimes where the 1-form is connected 
to the cross-terms between space and time components. 
Gibbons \textit{et al} \cite{ghww} studied the Zermelo 
navigation problem and connected it to Randers 
geometry via the optical metric for stationary spacetimes, 
and later Gibbons and Warnick applied the theory to study 
sound waves in the wind \cite{gw}. 
Very recently, the Zermelo formulation was used by 
Li and Jia to formulate new spacetimes \cite{lj} using 
radial and rotating vector fields.

The frame dragging effect was derived in 1918 by Lense 
and Thirring, known as the Lense Thirring effect 
\cite{thirring1, thirlens, thirring2}. 
It is a phenomenon describing the distortion of spacetime 
manifesting as a moving frame of reference upon which 
all motion takes place, originally discussed for massive 
rotating objects such as the Kerr Black Hole that involve 
stationary metrics. 
In \cite{epstein}, Epstein described a Hamiltonian 
mechanics approach to study frame dragging, demonstrating 
how it manifests with motion occurring despite the absence 
of momentum, differentiating between hidden and maifest 
frame dragging by discussing examples such as the expanding 
universe given by Friedmann-Robertson-Walker (FRW) metric, 
frames under constant acceleration, and spinning black holes 
such as the Kerr metric. 
In the first two examples, he showed how frame dragging 
is hidden in the Robertson-Walker metric and the constantly 
accelerated frames, manifesting only under a suitable 
co-ordinate transformation. 
In the case of Kerr metric, the frame dragging was made 
manifest to begin with. 
More recently Jaranowski, Mach, Malech, and Pir\'{o}g 
incorporated an anti-dragging effect into new relativistic 
extensions of Newtonian rotation laws \cite{jmmp, mm}. 
In their example they conisdered an accretion disk 
spinning around a Schwarzschild Black Hole examined 
using the Post-Newtonian Approximation.

Recently in \cite{chanda}, I had introduced a formulation 
of relativistic Hamiltonian mechanics on curved spaces 
of Randers' Finsler form using the momentum constraint 
as a generator for Hamiltonian equations of motion. 
Under inspiration from Epstein's work \cite{epstein}, 
I used my constraint mechanics to study frame dragging, 
and described frame dragging in the context of Jacobi 
metric, and thus, for Randers-Finsler metrics. 
Since Epstein's work can be compared to Gibbons' 
work on the Zermelo Navigation problem \cite{ghww}, 
it presents the opportunity to further generalise the 
problem for RF metrics, and include time-dependent 
Zermelo winds. 
The examples of expanding universe and accelerated 
frames in special relativity considered by Epstein 
present an opportunity to formulate and consider 
examples of such time-dependent winds.

In this article, we will describe the frame dragging 
effect in the context of the Zermelo navigation problem. 
After a brief review of required preliminaries, we will start 
by showing that the Zermelo wind is the frame dragging 
drift velocity using a Hamiltonian mechanics approach. 
There we discuss the afforementioned examples 
of the expanding universe and the constantly accelerating 
frame, casting them into Zermelo form and describing 
time-dependent Zermelo winds. 
I furthermore attempt to account for the anti-dragging 
effect in the Zermelo description of spacetime by 
describing the self-gravitating disk around a black hole 
discussed by Jaranowski {\it et al}.

\section{Preliminiaries} 

Here, I will review the prerequisite fundamentals 
necessary for the discussions to follow afterwards. 
First I will discuss Hamiltonian mechanics using 
the constraint on relativistic momentum. 
Then using constraint mechanics, I will describe 
a Hamiltonian mechanics approach to frame 
dragging along similar lines to Epstein's work.

\subsection{Hamiltonian mechanics with Constraint} 
\label{sec:rf} 

In RF metrics the first part with the norm under 
the square root accounts for the influence of 
curvature, while the linear term outside accounts 
for gauge field interaction. 
Sometimes the second part is geometric in origin. 
\begin{equation} \label{rfmetric} 
     ds = 
     \sqrt{g_{\mu \nu} (\bm x) d x^\mu d x^\nu} + 
     A_\lambda (\bm x) d x^\lambda. 
\end{equation} 
Using the Lagrangian $L$ and the canonical 
momenta $\bm p$ leads us to the gauge-covariant 
momenta $\bm \pi$ given by: 
$$
     L = 
     \sqrt{
          g_{\mu \nu} (\bm x) 
          \dot x^\mu \dot x^\nu
     } + 
     A_\lambda (\bm x) 
     \dot x^\lambda 
\qquad \Rightarrow \qquad 
     p_\mu = 
     \frac{\partial L}{\partial \dot x^\mu} = 
     \frac{g_{\mu \nu} (\bm x) \dot x^\nu}
     {\sqrt{g_{\alpha \beta} (\bm x) \dot x^\alpha \dot x^\beta}} + 
     A_\mu (\bm x),
$$
\begin{equation} \label{rfmom} 
     \pi_\mu = 
     p_\mu - A_\mu (\bm x) = 
     g_{\mu \nu} (\bm x) 
     \frac{d x^\nu}{d \sigma}, 
\qquad \text{where } \ 
     d \sigma := 
     \sqrt{
          g_{\alpha \beta} (\bm x) 
          d x^\alpha d x^\beta
     }.
\end{equation} 
Thus, an alternative generator of equations 
in phase space is given by the constraint 
obeyed by the gauge-covariant momenta 
$\bm \pi$ \eqref{rfmom}:
\begin{equation} \label{constraint} 
     \phi (\bm x, \bm p) = 
     \sqrt{
          g^{\mu \nu} (\bm x) 
          \pi_\mu \pi_\nu
     } = 
     \sqrt{
          g_{\mu \nu} (\bm x) 
          \frac{d x^\mu}{d \sigma} 
          \frac{d x^\nu}{d \sigma}} = 1,
\end{equation} 
which acts as a generator of equations of motion, 
demonstrated by taking a derivative of the constraint:
\begin{equation} \label{cons2} 
     \frac{d \phi}{d \sigma} = 
     \frac{\partial \phi}{\partial x^\mu} 
     \frac{d x^\mu}{d \sigma} + 
     \frac{\partial \phi}{\partial p_\mu} 
     \frac{d p_\mu}{d \sigma} = 0,
\end{equation}
then I can show by applying \eqref{rfmom} and \eqref{constraint} 
into \eqref{cons2} that one will have:
$$\frac{\partial \phi}{\partial p_\mu} = 
g^{\mu \nu} (\bm x) \pi_\nu = 
\frac{d x^\mu}{d \sigma}
\qquad \Rightarrow \qquad 
\frac{\partial \phi}{\partial x^\mu} = 
- \frac{d p_\mu}{d \sigma}.$$
Thus, we have the constraint equivalent of Hamilton's equations 
of motion:
\begin{equation} \label{conseq} 
     \boxed{
          \frac{d x^\mu}{d \sigma} = 
          \frac{\partial \phi}{\partial p_\mu} 
     \qquad , \qquad 
          \frac{d p_\mu}{d \sigma} = 
          - \frac{\partial \phi}{\partial x^\mu}
     }.
\end{equation}

\subsection{Frame dragging effect} 
\label{sec:frame_drag} 

According to Epstein \cite{epstein}, the frame dragging 
effect describes motion independent of momentum. 
If we have the stationary spacetime metric: 
\begin{equation} \label{stationary} 
     d s^2 = 
     g_{ij} (\bm x) 
     d x^i d x^j + 
     2 g_{i0} (\bm x) 
     d x^i d t + 
     g_{00} (\bm x) 
     d t^2,
\end{equation} 
then the constraint is given by: 
\begin{align} \label{statcons} 
     \phi (\bm{x, p}) &= 
     \sqrt{
          g^{ij} (\bm x) 
          p_i p_j + 
          2 g^{i0} (\bm x) 
          p_i p_0 + 
          g^{00} (\bm x) 
          (p_0)^2 
     } 
     \nonumber 
\\ 
     &= 
     \sqrt{
          f^{ij} (\bm x) 
          p_i p_j + 
          g^{00} (\bm x) 
          \left( 
               p_0 + 
               \frac{g^{0m} (\bm x)}{g^{00} (\bm x)} 
               p_m 
          \right)^2 
     } = 1, 
\qquad \text{where } 
     f^{ij} (\bm x) = 
     g^{ij} (\bm x) - 
     \frac{g^{i0} (\bm x) g^{j0} (\bm x)}{g^{00} (\bm x)}, 
\end{align} 
leading to the following constraint equations: 
\begin{equation} \label{statconseq} 
     \begin{split} 
          \frac{d x^i}{d s} &= 
          \frac{\partial \phi}{\partial p_i} = 
          g^{ij} (\bm x) p_j + 
          g^{i0} (\bm x) p_0 = 
          f^{ij} (\bm x) p_j + 
          \frac{g^{i0} (\bm x)}{g^{00} (\bm x)} 
          \left( 
               g^{00} (\bm x) 
               p_0 + 
               g^{0j} (\bm x) 
               p_j
          \right) 
     \\ 
          \frac{d t}{d s} &= 
          \frac{\partial \phi}{\partial p_0} = 
          g^{0j} (\bm x) p_j + 
          g^{00} (\bm x) p_0 = 
          \sqrt{
               g^{00} (\bm x) 
               \left( 
                    1 - f^{ij} (\bm x) 
                    p_i p_j 
               \right) 
          }
     \end{split}  
\end{equation} 
from which we can see that 
\begin{equation} \label{vel} 
     \frac{d x^i}{d t} = 
     \frac{f^{ij} (\bm x)}{1 - \sqrt{g^{00} (\bm x) \left( 1 - f^{ij} (\bm x) p_i p_j \right)}} 
     p_j + 
     \frac{g^{i0} (\bm x)}{g^{00} (\bm x)}, 
\end{equation} 
which matches what Epstein discussed in \cite{epstein}. 
Ultimately, frame dragging is manifested as motion 
or velocity that exists in the absence of momentum. 
This can also be seen from \eqref{vel} or directly from 
the constraint equations \eqref{statconseq}: 
\begin{equation} \label{fdrag} 
     \begin{split} 
          p_j = 0 \quad \forall \ j 
     \end{split} 
\qquad \Rightarrow \qquad 
     \left\{ \begin{split} 
          \left( 
               \frac{d x^i}{d s} 
          \right)_{p_j = 0} &= 
          g^{i0} (\bm x) 
          p_0 
     \\ 
          \left( 
               \frac{d t}{d s} 
          \right)_{p_j = 0} &= 
          g^{00} (\bm x) 
          p_0    
     \end{split} \right\} 
\qquad \Rightarrow \qquad 
     \left( 
          \frac{d x^i}{d t} 
     \right)_{p_j = 0} = 
     \frac{g^{i0} (\bm x)}{g^{00} (\bm x)} 
\end{equation}

\section{Zermelo form of metric} 
\label{sec:zermelo} 

We can rewrite the metric for a stationary spacetime 
\eqref{stationary} into another form given as: 
\begin{align} 
     d s^2 &= 
     g_{ij} (\bm x) 
     d x^i d x^j - 
     2 \left( 
          g_{ij} (\bm x) 
          \frac{g^{j0} (\bm x)}{g^{00} (\bm x)} 
     \right) 
          d x^i d t + 
     g_{00} (\bm x) 
     d t^2 
     \nonumber 
\\ 
     \label{prezermelo}
     &= 
     g_{ij} (\bm x) 
     \left( 
          d x^i - 
          W^i (\bm x) d t 
     \right) 
     \left( 
          d x^j - 
          W^j (\bm x) d t 
     \right) + 
     \left( 
          g_{00} (\bm x) - 
          g_{ab} (\bm x) 
          W^a (\bm x) W^b (\bm x) 
     \right) 
     d t^2, 
\end{align}  
where we have written a wind function $W^i (\bm x)$ as: 
\begin{equation} \label{wind} 
     W^i (\bm x) =  
     \frac{g^{i0} (\bm x)}{g^{00} (\bm x)} 
\qquad , \qquad 
     W_i (\bm x) = 
     g_{ij} (\bm x) 
     W^j (\bm x) = 
     - g_{i0} (\bm x). 
\end{equation} 
If we define 
$$
     g_{ij} (\bm x) := 
     - h_{ij} (\bm x) 
     \left( 
          g_{00} (\bm x) - 
          g_{ab} (\bm x) 
          W^a (\bm x) W^b (\bm x) 
     \right) 
$$
then we can further write: 
$$
     - h_{ij} (\bm x) = 
     \frac{g_{ij} (\bm x)}{g_{00} (\bm x) - 
          g_{ab} (\bm x) 
          W^a (\bm x) W^b (\bm x)}
$$ 
$$
\Rightarrow \qquad 
     - h_{ij} (\bm x) 
     W^i (\bm x) 
     W^j (\bm x) = 
     \frac{g_{ab} (\bm x) 
          W^a (\bm x) W^b (\bm x)}{g_{00} (\bm x) - 
          g_{ab} (\bm x) 
          W^a (\bm x) W^b (\bm x)}
$$ 
$$
\Rightarrow \qquad 
     1 - h_{ij} (\bm x) 
     W^i (\bm x) 
     W^j (\bm x) = 
     \frac{g_{00} (\bm x)}{g_{00} (\bm x) - 
          g_{ab} (\bm x) 
          W^a (\bm x) W^b (\bm x)}
$$ 
$$
\Rightarrow \qquad 
     g_{00} (\bm x) - 
     g_{ab} (\bm x) 
     W^a (\bm x) W^b (\bm x) = 
     \frac{g_{00} (\bm x)}{1 - h_{ij} (\bm x) 
     W^i (\bm x) 
     W^j (\bm x)}
$$ 
Thus, we can write \eqref{prezermelo} as: 
$$ 
     d s^2 = 
     \frac{g_{00} (\bm x)}{1 - h_{ij} (\bm x) 
     W^i (\bm x) 
     W^j (\bm x)} 
     \left[ 
          d t^2 - 
          h_{ij} (\bm x) 
          \left( 
               d x^i - 
               W^i (\bm x) d t 
          \right) 
          \left( 
               d x^j - 
               W^j (\bm x) d t 
          \right) 
     \right], 
$$  
Now if we write the stationary metric \eqref{stationary} as: 
\begin{equation} \label{station} 
     d s^2 = 
     - \left( V (\bm x) \right)^2 
     \left( 
          d t + 
          \omega_i (\bm x) 
          d x^i 
     \right)^2 + 
     \gamma_{ij} (\bm x) 
     d x^i 
     d x^j, 
\end{equation} 
where:  
\begin{equation} \label{legend1}
     \left( V (\bm x) \right)^2 = 
     - g_{00} (\bm x) 
\qquad , \qquad 
     \omega_i (\bm x) = 
     - \frac{g_{i0} (\bm x)}{g_{00} (\bm x)}
\end{equation}  
\begin{equation} \label{legend2} 
     W^i (\bm x) = 
     \left( V (\bm x) \right)^2 
     \gamma^{ij} (\bm x) 
     \omega_j (\bm x) 
\qquad \Rightarrow \qquad 
     \omega_i (\bm x) = 
     \frac{1}{\left( V (\bm x) \right)^2} 
     g_{ij} (\bm x) 
     W^j (\bm x)
\end{equation} 
\begin{equation} \label{legend3} 
\boxed{
     h_{ij} (\bm x) = 
     \frac{1 - \left(V (\bm x) \right)^2 g^{ab} (\bm x) \omega_a (\bm x) \omega_b (\bm x)}{\left(V (\bm x) \right)^2} 
     g_{ij} (\bm x),  
}
\end{equation} 
then we will have the Zermelo form of the metric 
discussed by Gibbons, Herdeiro, Warnick, and 
Werner \cite{ghww}: 
\begin{equation} \label{zermelo} 
\boxed{
     d s^2 = 
     \frac{\left(V (\bm x) \right)^2}{1 - h_{ij} (\bm x) W^i (\bm x) W^j (\bm x)} 
     \left[ 
          - d t^2 + 
          h_{ij} (\bm x) 
          \left( 
               d x^i - W^i (\bm x) d t 
          \right) 
          \left( 
               d x^j - W^j (\bm x) d t 
          \right) 
     \right], 
}
\end{equation} 
I will now consider an example of the Kerr metric 
for application of the theory. 
\bigskip

The Kerr metric is the first and simplest example 
one can consider of a spinning black hole used 
when discussing the frame dragging effect around 
rotating masses. 
The Kerr metric describes a rotating uncharged 
black hole that is a generalisation of the Schwarzschild 
black hole to include rotation, the exact solution of which 
was discovered by Kerr in 1963 \cite{kerr}. 
\bigskip 

\noindent 
In Boyer-Lindquist co-ordinates the Kerr metric 
is given by: 
\begin{equation} \label{kerrmetric} 
     d s^2_R = 
     \left( 
          1 - \frac{2 M r}{\rho^2} 
     \right) 
     d t^2 + 
     \frac{4 M a r \sin^2 \theta}{\rho^2} 
     d t \; d \varphi - 
     \rho^2 
     \left[ 
          \frac{dr^2}{\Delta} + d \theta^2 + 
          \frac{\sin^2 \theta}{\rho^4} 
          \left\{ 
               \left( 
                     r^2 + a^2 
               \right)^2 - 
               a^2 \Delta \sin^2 \theta 
          \right\} 
          d \varphi^2 
     \right], 
\end{equation} 
where $\Delta (r) = r^2 - 2 M r + a^2$ , and 
$\rho^2 (r, \theta) = r^2 + a^2 \cos^2 \theta$. 
Writing \eqref{kerrmetric} in Zermelo form 
according to \eqref{legend1}, \eqref{legend2}, 
and \eqref{legend3} gives us: 
\begin{equation} \label{kerrzer} 
     \left(V (\bm x) \right)^2 = 
     - \left( 
          1 - \frac{2 M r}{\rho^2} 
     \right) 
\qquad , \qquad 
     W^\varphi (\bm x) = 
     \frac{2 M a r} 
     {
          \left( 
                r^2 + a^2 
          \right)^2 - 
          a^2 
          \Delta 
          \sin^2 \theta 
     }, 
\end{equation} 
Naturally, the Zermelo wind gradually disappears 
as one goes far away from the spinning black hole.

\subsection{Frame dragging in Zermelo form} 
\label{sec:zerfdrag}

From the Zermelo form \eqref{zermelo} we have the constraint: 
\begin{equation} \label{zercons} 
     \phi (\bm{x, p}) = 
     \sqrt{ 
          \frac{1 - h_{ij} (\bm x) W^i (\bm x) W^j (\bm x)}{\left(V (\bm x) \right)^2} 
          \left[ 
               \left( 
                    p_0 + 
                    W^i (\bm x) 
                    p_i 
               \right)^2 - 
               h^{ij} (\bm x) 
               p_i p_j 
          \right] 
     } = 1,
\end{equation} 
where we have: 
\begin{equation} \label{legend4} 
\boxed{
     h^{ij} (\bm x) = 
     \frac{\left(V (\bm x) \right)^2} 
     { 
          1 - 
          \left(V (\bm x) \right)^2 
          g^{ab} (\bm x) 
          \omega_a (\bm x) 
          \omega_b (\bm x) 
     }  
     \gamma^{ij} (\bm x). 
}
\end{equation} 
So when we take the constraint equations \eqref{conseq} 
we shall have: 
\begin{equation} \label{zervel} 
     \begin{split} 
          \frac{d x^i}{d s} &= 
          \frac{1 - h_{ij} (\bm x) W^i (\bm x) W^j (\bm x)}{\left(V (\bm x) \right)^2} 
          \left[ 
               - h^{ij} (\bm x) 
               p_j + 
               W^i (\bm x) 
               \left( 
                    p_0 + 
                    W^a (\bm x) 
                    p_a 
               \right)
          \right]  
     \\ 
          \frac{d t}{d s} &= 
          \frac{1 - h_{ij} (\bm x) W^i (\bm x) W^j (\bm x)}{\left(V (\bm x) \right)^2} 
          \left( 
               p_0 + 
               W^i (\bm x) 
               p_i 
          \right) 
     \end{split}, 
\end{equation} 
from which we will have the frame dragging velocity: 
\begin{equation} \label{zerfdrag} 
     \left( 
          \frac{d x^i}{d t} 
     \right)_{p_j = 0} = 
     W^i (\bm x). 
\end{equation} 
Thus, the frame dragging velocity can be shown 
to be the drift with the Zermelo wind. 
\bigskip 

Previously, in \cite{chanda}, I discussed the frame 
dragging in the context of the Jacobi metrics of 
stationary spacetimes, which have a RF form. 
This naturally begs the question of how to discuss 
the Zermelo navigation problem for Jacobi's metric 
and RF spacetimes.
Here, I will try to connect the Zermelo form to the 
Jacobi-Maupertuis-Randers-Finsler (JMRF) metric. 
Now we know that the Jacobi metric for a stationary 
spacetime is given by: 
\begin{equation} \label{rfjmet} 
     d s_J = 
     \sqrt{
          \left( 
               1 - \frac{q^2}{g_{00} (\bm x)} 
          \right) 
          \gamma_{ij} (\bm x) d x^i d x^j
     } + 
     q
     \frac{g_{0i} (\bm x)}{g_{00} (\bm x)} 
     d x^i.
\end{equation} 
In Zermelo form, this metric can be written as: 
\begin{equation} \label{zerjmet2} 
     d s_J = 
     \sqrt{
          \left( 1 - \frac{q^2}{\left( V (\bm x) \right)^2} \right) 
          \left( 
               \frac{\left(V (\bm x) \right)^2}{1 - \left(V (\bm x) \right)^2 g^{ab} (\bm x) \omega_a (\bm x) \omega_b (\bm x)} 
               h_{ij} (\bm x) + 
               \frac{W_i (\bm x) W_j (\bm x)}{\left( V (\bm x) \right)^2} 
          \right) 
          d x^i d x^j
     } + 
     q
     \frac{W_i (\bm x)}{\left( V (\bm x) \right)^2} 
     d x^i.
\end{equation} 
Having described the Jacobi metric for stationary 
metrics in terms of Zermelo data, we should now 
consider the Jacobi metric for a RF metrics and 
how a Zermelo wind could be described from 
RF metrics.

\subsection{Zermelo wind for Randers-Finsler metrics} 
\label{sec:zermelorf}
 
Consider the following RF metric derived by adding 
a linear 1-form term to \eqref{zermelo}: 
\begin{equation} \label{zrf} 
     d \widetilde s = 
     \sqrt{
          \frac{\left(V (\bm x) \right)^2}{1 - h_{ij} (\bm x) W^i (\bm x) W^j (\bm x)} 
          \left[ 
               - d t^2 + 
               h_{ij} (\bm x) 
               \left( 
                    d x^i - W^i (\bm x) d t 
               \right) 
               \left( 
                    d x^j - W^j (\bm x) d t 
               \right) 
          \right]
     } + 
     A_i (\bm x) 
     d x^i, 
\end{equation} 
for which we will have the constraint: 
\begin{equation} \label{zrfcons} 
     \phi (\bm{x, p}) = 
     \sqrt{ 
          \frac{1 - h_{ij} (\bm x) W^i (\bm x) W^j (\bm x)}{\left(V (\bm x) \right)^2} 
          \left[ 
               \left( 
                    p_0 + 
                    W^i (\bm x) 
                    \pi_i 
               \right)^2 - 
               h^{ij} (\bm x) 
               \pi_i \pi_j 
          \right] 
     } = 1, 
\qquad \text{where } \ 
     \pi_i = p_i - A_i (\bm x).
\end{equation} 
This time if we take the constraint equations \eqref{conseq} 
we shall have: 
\begin{equation} \label{zrfvel} 
     \begin{split} 
          \frac{d x^i}{d s} &= 
          \frac{1 - h_{ij} (\bm x) W^i (\bm x) W^j (\bm x)}{\left(V (\bm x) \right)^2} 
          \left[ 
               - h^{ij} (\bm x) 
               \pi_j + 
               W^i (\bm x) 
               \left( 
                    p_0 + 
                    W^a (\bm x) 
                    \pi_a 
               \right) 
          \right]  
     \\ 
          \frac{d t}{d s} &= 
          \frac{1 - h_{ij} (\bm x) W^i (\bm x) W^j (\bm x)}{\left(V (\bm x) \right)^2} 
          \left( 
               p_0 + 
               W^a (\bm x) 
               \pi_a 
          \right) 
     \end{split}, 
\end{equation} 
from which we will have the frame dragging velocity: 
\begin{equation} \label{zerfdrag} 
     \widetilde W^i (\bm x) = 
     \left( 
          \frac{d x^i}{d t} 
     \right)_{p_j = 0} = 
     W^i (\bm x) + 
     \frac{h^{ij} (\bm x) A_j (\bm x)}{p_0 - W^a (\bm x) A_a (\bm x)}. 
\end{equation} 
Thus, the gauge field term acts to contribute an additional 
component to the Zermelo wind observed.

\section{Time dependent Zermelo winds} 
\label{sec:tdwind}

In \cite{ghww}, Gibbons {\it et al} discussed 
generalising the Zermelo navigation problem 
by including time-dependent winds in the 
formulation. 
Since I have described Zermelo winds as 
a geometrization of the frame-dragging 
effect, I have included two examples of 
time-dependent spacetimes considered 
by Epstein in his study of frame dragging 
\cite{epstein}.

\subsection{Expanding Universe - FRW spacetime}
\label{sec:frw} 

The Friedmann-Robertson-Walker metric is an exact 
solution of Einstein field equations in general relativity 
that is homogeneous, isotropic, and expanding \cite{wald}. 
This was one of 2 examples discussed by Epstein \cite{epstein} 
to describe frame dragging in the expanding universe. 
\bigskip 

\noindent 
The expanding universe metric is given by the 
spatially flat FRW metric: 
\begin{equation} \label{frw1} 
     d s^2 = 
     d t^2 - 
     \left( a (t) \right)^2 
     \delta_{ij} 
     d X^i d X^j, 
\end{equation} 
where $a (t)$ is the time-dependent scale factor 
and $X^i, \ i = 1, 2, 3$ are the Cartesian co-ordinates. 
Since the metric involves time dependence, 
it is not possible to formulate a corresponding 
Jacobi metric. 
Writing the co-ordinates as $X^i = \frac{x^i}{a (t)}$, 
we can rewrite the FRW metric \eqref{frw1} as: 
\begin{equation} \label{frw2} 
     d s^2 = 
     \left[ 
          1 - 
          \left( Q (t) \right)^2 
          r^2
     \right] 
     d t^2 - 
     2 Q (t) 
     \delta_{ij} x^i \; 
     d t d x^j - 
     \left( 
          \delta_{ij} 
          d x^i d x^j 
     \right), 
\end{equation} 
$$
     \text{where } \ 
     r^2 = \delta_{ij} x^i x^j 
\ , \ 
     Q (t) = \frac{a' (t)}{a (t)}, 
$$
$Q (t)$ being the Hubble factor. 
Thus, if we were to write \eqref{frw2} in Zermelo 
form \eqref{zermelo}, then we will have according 
to \eqref{legend1}, \eqref{legend2}, and \eqref{legend3}: 
\begin{equation} \label{frwzer} 
\begin{split} 
     \left(V (\bm x) \right)^2 &= 
     - \left[ 
          1 - 
          \left( Q (t) \right)^2 
          r^2
     \right], 
\\ 
     W^i (\bm x) &= 
     Q (t) 
     x^i, 
\\      
     h_{ij} (\bm{x}, t) &= 
     \frac{1}{\left(1 - \left( Q (t) \right)^2 r^2 \right) + \left( Q (t) \right)^2 r^2} 
     \delta_{ij} 
\end{split} 
\end{equation} 
showing that the Zermelo wind is time-varying 
and radially linear from the origin.

\subsection{Accelerated frames in Special Relativity}
\label{sec:specrel}

Consider the Lorentzian line element in Cartesian 
co-ordinates in the frame of constant acceleration 
$a$ in $1 + 1$ spacetime: 
\begin{equation} \label{constacc1} 
     d s^2 = 
     d T^2 - d X^2, 
\qquad \text{where } \ 
     X = x + \frac12 a t^2
\quad , \quad  
     T = t + a x t. 
\end{equation} 
This metric can be rewritten as: 
\begin{equation} \label{constacc2} 
     d s^2 = 
     \left( 
          1 + 
          2 a x + 
          a^2 x^2 - 
          a^2 t^2 
     \right) 
     d t^2 + 
     2 a^2 t x \; 
     d t d x - 
     \left( 
          1 - a^2 t^2 
     \right) 
     d x^2. 
\end{equation} 
The constraint for this metric is given by: 
\begin{equation} \label{constracc} 
     \phi (\bm{x, p}) = 
     \sqrt{ 
          - \frac{1 + 2 a x + a^2 x^2 - a^2 t^2}{\left( 1 + a x + a^2 t^2 \right)^2} 
          \left( 
               p_x - 
               \frac{a^2 t x}{1 + 2 a x + a^2 x^2 - a^2 t^2} 
               p_t 
          \right)^2 + 
          \frac{\left( p_t \right)^2}{1 + 2 a x + a^2 x^2 - a^2 t^2}
     } = 1
\end{equation} 
Naturally, in Zermelo form according to \eqref{legend1}, 
\eqref{legend2}, and \eqref{legend3}, we would have: 
\begin{equation} \label{acczer} 
\begin{split} 
     \left(V (\bm x) \right)^2 &= 
     - \left( 
          1 + 
          2 a x + 
          a^2 x^2 - 
          a^2 t^2
     \right), 
\\ 
     W^x (\bm x) &= 
     \frac{\left( 1 + a x \right)^2 - a^2 t^2}{\left( 1 + a x + a^2 t^2 \right)^2 - a^4 t^2 x^2} 
     a^2 t x, 
\\ 
     h_{xx} (x, t) &= 
     \frac{1 - a^2 t^2}{\left( 1 + a x \right)^2 - a^2 t^2 + \left(1 - a^2 t^2 \right) \left( W^x \right)^2} 
\end{split} 
\end{equation}

\section{Self-gravitating accretion disk} 
\label{sec:disk} 

In 2015, Jaranowski, Mach, Malech, and Pir\'{o}g 
\cite{jmmp} studied a Schwarzschild black hole 
with a self-gravitating accretion disk resulting from 
a binary merger of a black hole with a neutron star. 
By applying the 1st Post-Newtonian (1PN) 
approximation, they discovered an anti-dragging 
effect. 
\bigskip 

\noindent 
The stationary metric for the 1st Post 
Newtonian (1PN) approximation: 
\begin{equation} \label{1pndisk1} 
     d s^2 = 
     \left( 
          1 + 
          2 \frac{U (r)}{c^2} + 
          2 \frac{\left( U (r) \right)^2}{c^4} 
     \right) 
     c^2 d t^2 + 
     2 \frac{A_i (r)}{c^2} 
     dt \; d x^i - 
     \left( 
          1 - 
          2 \frac{U (r)}{c^2} 
     \right) 
     \left( 
          \delta_{ij} 
          d x^i d x^j  
     \right) 
\end{equation} 
Accounting for azimuthal and equatorial symmetry, 
we can rewrite the metric into: 
\begin{equation} \label{1pndisk2} 
     d s^2 = 
     \left( 
          1 + 
          2 \frac{U (r)}{c^2} + 
          2 \frac{\left( U (r) \right)^2}{c^4} 
     \right) 
     c^2 d t^2 + 
     2 \frac{A_\varphi (r)}{c^2} 
     dt \; d \varphi - 
     \left( 
          1 - 
          2 \frac{U (r)}{c^2} 
     \right) 
     \left( 
          d r^2 + 
          r^2 
          d \Omega^2 
     \right) 
\end{equation} 
where $A_\varphi (r) \geq 0$. 
Here, in Zermelo form according to \eqref{legend1}, 
\eqref{legend2}, and \eqref{legend3}, we would have: 
\begin{equation} \label{1pnzer} 
\begin{split} 
     \left(V (\bm x) \right)^2 &= 
     - \left( 
          1 + 
          2 \frac{U (r)}{c^2} + 
          2 \frac{\left( U (r) \right)^2}{c^4} 
     \right), 
\\  
     W^\varphi (\bm x) &= 
     \frac{1}{c^2 - 2 U (r)} 
     \delta^{\varphi j} 
     A_j (r) = 
     \frac{1}{c^2 - 2 U (r)} 
     A_\varphi (r) 
\\ 
     h_{ij} (\bm{x}) &= 
     \frac{1 - 2 \frac{U (r)}{c^2}}{\left(1 + 2 \frac{U (r)}{c^2} + 2 \frac{\left( U (r) \right)^2}{c^4} \right) c^2 + \left( 1 - 2 \frac{U (r)}{c^2} \right) \left| A (r) \right|^2} 
     \delta_{ij} 
\end{split}      
\end{equation} 
Now, if we consider the 1PN approximation 
discussed in \cite{jmmp}: 
\begin{equation} \label{approx1pn} 
     U (r) = 
     U_0 (r) + 
     \frac1{c^2} 
     U_1 (r), 
\end{equation} 
where $U_0$ is the Newtonian gravitational potential: 
$$
     U_0 (r) = 
     - \frac{G M_c}{R} + 
     U^D_0, 
\qquad 
     \Delta U^D_0 = 
     4 \pi G \rho_0, 
$$
then we will have the Zermelo wind given by \eqref{1pnzer}: 
\begin{equation} \label{wind1pn} 
     W^\varphi (r) \approx 
     \frac1{c^2} 
     A_\varphi (r) 
     \left( 
          1 + 
          \frac2{c^2} 
          U_0 (r) + 
          \frac2{c^4} 
          U_1 (r) 
     \right), 
\end{equation} 
where according to \cite{jmmp}, we have: 
\begin{equation} \label{1pndata} 
     \begin{split} 
          \Delta A_\varphi &- 
          2 \frac{\partial_r A_\varphi}{r} = 
          - 16 \pi G 
          \omega_0 
          \sqrt{r} \rho_0, 
     \\ 
           \Delta U_1 &= 
           4 \pi G 
           \left(
                M_c 
                U^D_0 (0) 
                \delta (\bm x) + 
                \rho_1 + 
                2 K \rho^{\frac53}_0 + 
                \rho_0 
                \left( 
                     h_0 - 
                     2 U_0 + 
                     2 \omega^2_0 
                     r^{-1}
                \right) 
           \right) 
      \\ 
           h_1 &= 
           - U_1 - 
           A_\varphi 
           \omega_0 
           r^{-\frac32} + 
           2 h_0 
           \omega^2_0 
           r^{-1} + 
           \frac12 
           \omega^4_0 
           r^{-2} - 
           \frac32 
           h_0 - 
           4 h_0 
           U_0 - 
           2 (U_0)^2 - 
           C_1, 
     \end{split} 
\end{equation} 
which upon solving gives the functions necessary 
in \eqref{wind1pn}.

\section{Conclusion and Discussion}

In this article, I have demonstrated that 
the Zermelo wind from Zermelo's navigation 
problem is the drift velocity resulting from 
the frame dragging effect using a Hamiltonian 
mechanics approach with the constraint. 
Furthermore, I have described Zermelo winds 
for RF spacetimes, showing that the additive 
Finsler term contributes an additional component. 

Afterwards, I have recast the examples 
discussed in Epstein's study of frame 
dragging to provide the Zermelo data. 
These two examples present two examples 
of Zermelo problems with a time-dependent 
wind. 
I have also included the self-gravitating disk 
in the Post-Newtonian approximation discussed 
by Jaranowski {\it et al} \cite{jmmp} into Zermelo 
form.

\section*{Acknowledgement} 

I wish to acknowledge Prof. Sanved Kolekar for helpful 
advice and discussions that led to the preparation of 
this article.

\end{document}